\documentclass[journal,transmag]{IEEEtran}
\usepackage{cite}
\usepackage{amsmath}
\usepackage{amssymb}
\usepackage{epsfig}

\def\vx{{\mathbf x}}

\def\vj{{\mathbf j}}

\def\vf{{\mathbf f}}

\def\vX{{\mathbf X}}

\begin{document}
\title{Feature Analysis for Classification of Physical Actions using surface EMG
Data}
\author{{Anish C. Turlapaty\IEEEauthorrefmark{1} ~\IEEEmembership{Member,~IEEE}
and Balakrishna
Gokaraju\IEEEauthorrefmark{2},~\IEEEmembership{Member,~IEEE}}
\thanks{ {\IEEEauthorrefmark{1} Indian Institute of Information
Technology Sri City, 	Sri City, AP 517646, India.} {\IEEEauthorrefmark{2}The
University of West Alabama,
 Department of CIST,
  Livingston, AL 35470, USA.}
 Corresponding author: A. Turlapaty email: anish.turlapaty@gmail.com}}
\maketitle
\markboth{Submitted to Journal of Biomedical and Health Informatics}%
{Turlapapty and Gokaraju: Feature Analysis for Classification of Physical Actions
using surface EMG Data}

\begin{abstract}
Based on recent health statistics, there are several thousands of people  with
limb disability and gait disorders that require a medical assistance. A robot
assisted rehabilitation therapy can help them recover and return to a normal life.
In this scenario, a successful methodology is to use the EMG signal based
information to control the support robotics. For this mechanism to function
properly, the EMG signal from the muscles has to be sensed and then the
biological motor intention has to be decoded and finally the resulting information
has to be communicated to the controller of the robot. An accurate detection of
the motor intention requires a pattern recognition based categorical
identification. Hence in this paper, we propose an improved classification
framework by identification of the relevant features that drive the pattern
recognition algorithm. Major contributions include a set of  modified spectral
moment  based features and another relevant inter-channel correlation feature
that contribute to an improved classification performance. Next, we conducted a
 sensitivity analysis of the classification algorithm to different EMG channels.
 Finally, the classifier performance is compared to that of the other state-of the
 art algorithms. \end{abstract}



%
%
\IEEEdisplaynontitleabstractindextext

\IEEEpeerreviewmaketitle


\section{Introduction}
\label{sec:intro}
\subsection{Background}
\IEEEPARstart{T}{he} physical disabilities have been a major problem in the
modern world due to various reasons. For example, the aging brings problems
such as the gait disorders and limb impairments leading to a loss of the quality of
life \cite{stolze2005prevalence}. Next the occupational, trauma, and sports based
injuries and other kind of severe accidents usually render the people either
completely or partially disabled. Another major cause of the disabilities in adults is
a stroke to the motor cortex or other related  regions. For example, an ischemic
stroke can lead to motor disabilities, most commonly to the upper limbs
\cite{Proietti2016}. In terms of the statistics, there are $300$ amputees per year
in the United Kingdom \cite{Al-Timemy2013} and a total of $185,000$ amputees
live in the United States \cite{Al-Timemy2016}. Most of them need a prosthetic
limb or a partial limb support. Only a few treatment options are available for
these individuals.  For instance, a therapeutic rehabilitation can assist the partially
disabled people for functional recovery in order to resume normal activities and
thus improve the quality of life. In this context, the wearable robots can assist in
improving the effectiveness of the rehabilitation \cite{lo2010robot}. There is a
growing evidence that a robot assisted therapy provides an improvement of the
motor skills \cite{JarrassA2014}, \cite{Grosu2015}. The key reason for this
improvement is due to the increased therapeutic repetitions and enhanced
motivation for the patient due to the involvement of the virtual reality and video
gaming \cite{Lum2002}. The wearable robotics can also be used to ease the
burden on the human beings in various manual tasks. For example, the disabled
can be supported by an externally powered prosthetic or an orthotic exoskeleton
to restore the limb functionality to some extent \cite{Benatti2017},
\cite{Kazerooni2007a}.

In most of these exoskeleton applications, the key goal is to build a human robot
interface that can learn the intention of the user and adapt itself in order to
provide an accurate and timely assistance \cite{rosen2001myosignal}. Note that
in a healthy human, there is an active communication between the central
nervous system via motor neurons with the concerned muscle groups to produce
the intended motion \cite{Shenoy2008}. However, in an amputee or in a person
with a stroke induced disability, the bioelectrical signal may not reach the muscles
or the muscles may not produce the required force. In these scenarios, an
engineering solution is to sense the required myoelectric signal, assuming the
signal has a causal relation to the intended motion, using an sEMG sensor on the
muscle surface and extract the relevant information. For example, one can
estimate the joint force and decode the intended direction needed for the
movement and relay it to the controller to generate the intended motion
\cite{Lenzi2012}, \cite{Khokhar2010}. The sEMG basically refers to the electrical
activity associated with the muscles when the muscle fibers are recruited via
neuronal firings \cite{DeLuca1979}.  A practical exoskeleton product based on
surface EMG was invented by an MIT based startup, Myomo, assists people with
partial disabilities to perform their daily activities \cite{VacaBenitez2013}.  In this
type of research, a major challenge is the design of suitable algorithms and the
hardware to cope up with the complexity of human systems and further adapt in
an uncertain environment \cite{Scheme2011}. In the recent years, there is a
growing evidence that the pattern recognition (PR) algorithms can be used to
learn the motor intention from the surface EMG signals\cite{Khezri2007}. In the
next subsection, we present a survey of the most recent algorithms for PR based
control of exoskeletons.
\subsection{Literature Review}
 This technology  usually has two components:
(1) an algorithm section that does relevant feature extraction and selection and
categorization of movement using a statistical discriminator and (2) a hardware
component to interface with an exoskeleton via a controller. In the literature, as
discussed below, we find a great deal of emphasis on the algorithm component
specifically on the feature extraction. In \cite{Shenoy2008}, an online method was
developed that has a PR stage and a control mechanism. The classification was
performed for $8$ hand gestures using the RMS value as a feature and a linear
SVM achieving $95$\% accuracy. Next an orthotic arm guided by an EMG based
control with upto four degrees of freedom was demonstrated for performing
basic manoeuvres in a 3D environment such as simple arm movements, an object
pick up and drop and finally a stacking task. In \cite{tsuji2010biomimetic}, an
ANN based control was developed that has sensitivity to the multiple levels of
movements of the flexors and the extensors to directly control the orthotic arm.
In \cite{Khokhar2010}, a PR algorithm was developed to determine the torque
level from a human wrist and use it for real time control of an exoskeleton
prototype. It was shown that a set of four EMG channels is sufficient for the
classification of different wrist actions at different torque levels. In
\cite{Al-Timemy2013}, a classification experiment was done, in which three
healthy subjects and three amputees participated. The healthy subjects were
asked to perform $15$ different finger movements and the amputees were asked
to imagine $12$ different movements. Using the frequency domain features and
the SVM, the movements were classified with an accuracy greater than $90$\%.
In \cite{Sapsanis2013}, the intrinsic mode functions were used for feature
extraction toward the classification of six categories of the hand grips. In another
work, the cardinality of the EMG signal was proposed as a highly relevant feature
for identifying the motor intent \cite{ortiz2015cardinality}.  Next, in
\cite{Al-Timemy2016}, a classification of six hand grips and finger positions was
analysed against the force levels. Nine amputees participated in the study and the
key innovation is introduction of the moment ratios as discriminative features for
a successful classification with an accuracy over $90$\% despite discrete
variations in the force levels. Next, \cite{Adewuyi2016} performed an analysis of
the relevant features and  the selection of corresponding classifiers for control of
the exoskeletons for partially disabled hands. They addressed the classification of
sEMG signals corresponding to four hand positions for $20$ subjects of which
$16$ are non-amputees and $4$ are partial hand amputees. They have compared
the efficiency of the time domain features, the time domain and autoregressive
features and the frequency domain features. Next in \cite{Ertugrul2015}, the local
binary pattern based features were used for the classification of the physical
action EMG data \cite{Lichman:2013}. A major drawback in \cite{Ertugrul2015}, is
that it is not clear whether the  analysis is for a binary classification  i.e, the
normal vs. the aggressive or for classification of all the $20$ categories of the
data. In the current paper, we address the above mentioned multi-category
classification problem as described in the following sections.

\subsection{Contributions}
In our paper, we address the classification of M-categories of physical actions
based on multi-channel EMG data. We propose an improved feature set
consisting of selected feature subsets from different feature modalities such as
the time domain (TD) statistics, the inter-channel TD statistics, the spectral
moments ratios and products, the spectral band powers and the local binary
pattern based statistics. In the feature extraction apart from other well known
features, we have modified the spectral moment features to improve the
classification performance. Moreover we have identified an inter-channel
correlation feature that also contributes to an improved classification. Next we
present the key EMG channels that significantly contribute toward the
performance improvement. Finally, we also demonstrate that the probabilistic
neural network classifier has similar performance to that of the kernel based SVM
classifier.
\section{Methodology}
\label{METHODS} 
\begin{figure*}
\onecolumn
  \centering
\includegraphics[width=16cm]{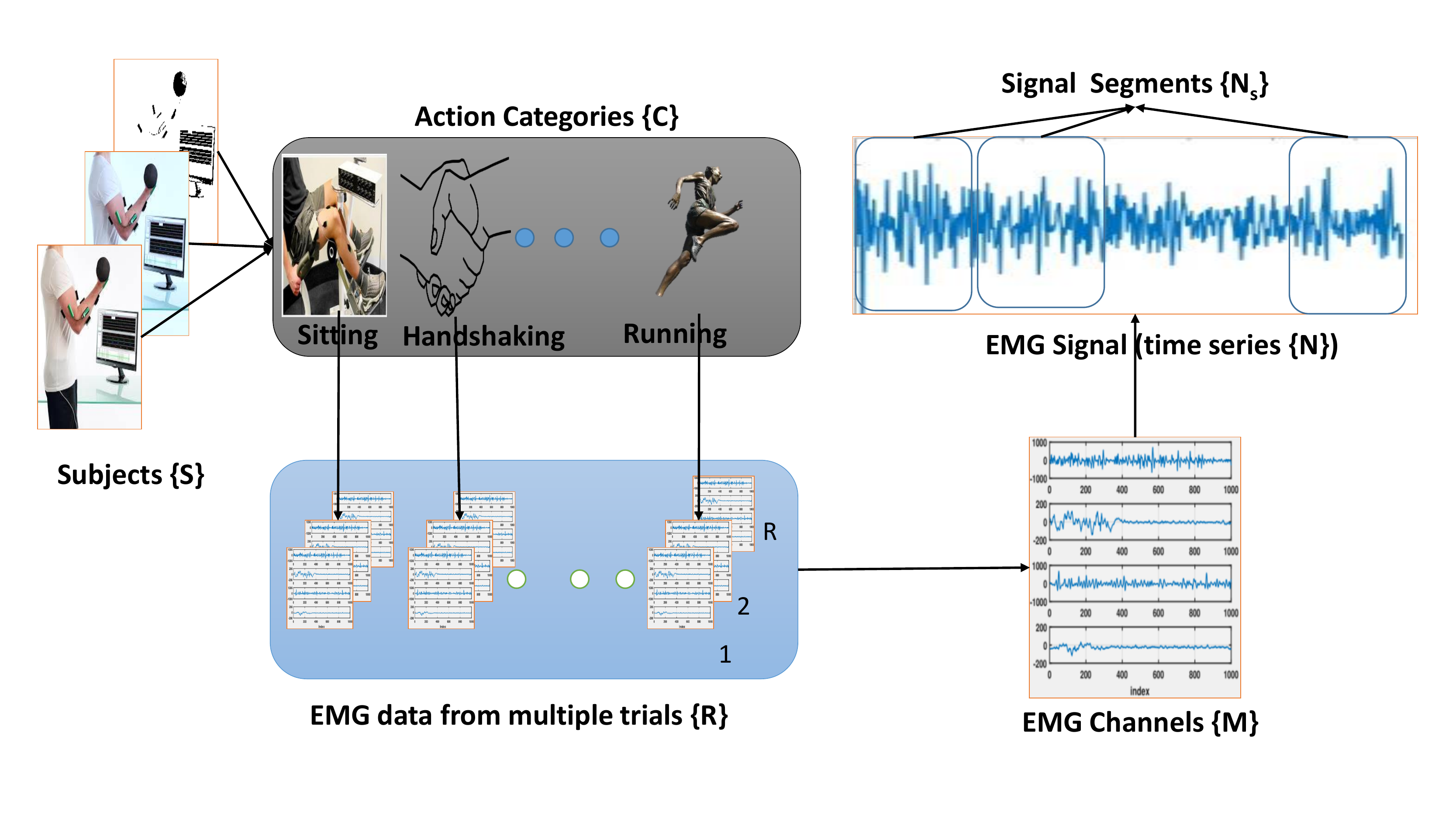}\\
\caption{Physical action data: setup for classification}\label{scheme1}
\end{figure*}

We consider a dataset $\vX$ with $P (= S \times C \times R)$
observation arrays obtained from $S$ subjects. These observations consist of
$C$ categories and $R$ trials each ie., each of the $S$ subjects have performed
each activity $R$ times as represented in fig. \ref{scheme1}. Finally, each of the
$p$-th observation array $X_p$ has $M$ channels (distinct EMG electrode
contact locations) and in each of the $m$-th channel $\vx_p^m$ there are $N$
values (samples).

\begin{eqnarray} \label{Dataset}
\vX &=& \bigg \{ X_1, X_2, \cdots , X_P \bigg \}   \\
X_p &=&   \bigg [\vx_p^{1}, \vx_p^{2}, \cdots, \vx_p^{M} \bigg ] ~~~ p =
1,\cdots,P  \nonumber
\\\vx_p^{m} &=& \bigg [x_p^m(1), x_p^m(2), \cdots, x_p^m(N) \bigg]^T
\nonumber \\ &&m = 1,\cdots, M\nonumber
\end{eqnarray}

The first step is segmentation of the EMG signal in each channel with a
non-overlapping and a sliding window length $L$, which is derived based on the
sampling rate of the concerned EMG data. The standard window length (in time)
is $200$ms from literature, see \cite{Khokhar2010} and \cite{Al-Timemy2013}.
Next, the $l$-th element of the $w$-th segment of the $m$-th channel of the
$p$-th observation is denoted as
\begin{equation} \label{Segmentation}
s_\vj(l) = x_{\vj}((w-1)*L+l)
\end{equation}
here, $\vj$ represents the index triplet $\vj = (m,w,p)$, where $w = 1, \cdots,
N_w$, $l = 1, \cdots, L$ and $N_w = \frac{N}{L}$. Note that each pattern can
consist of $N_s$ segments and features from each segment are combined to
build the complete feature vector corresponding to a pattern.

\begin{figure*}

  \centering
\includegraphics[width=16cm]{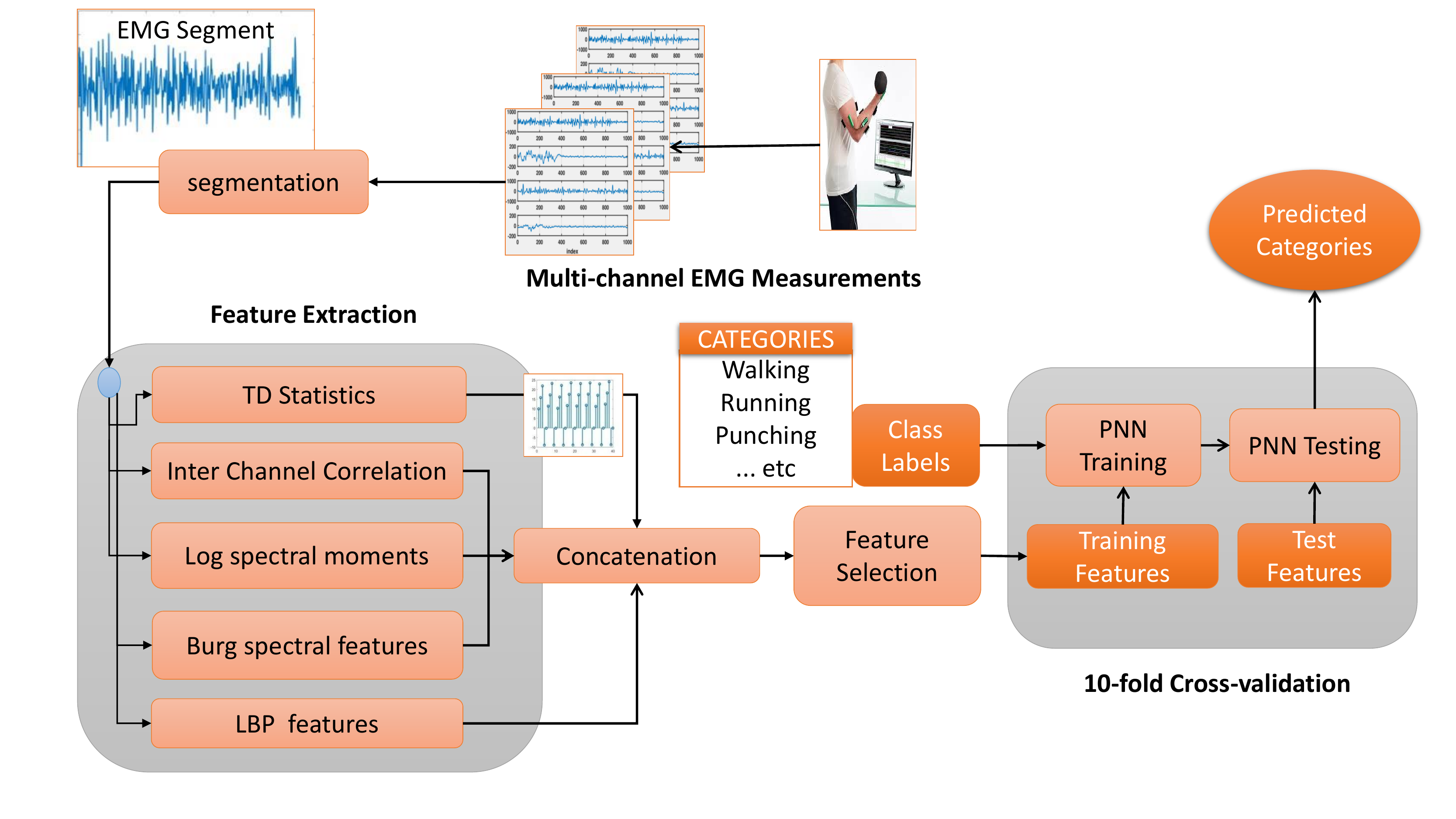}\\
\caption{Physical action data classification scheme using PNN}\label{scheme2}
\end{figure*}

Now we give an overview of the proposed pattern recognition framework as
given in fig. \ref{scheme2}. The EMG signals are segmented and the features from
various modalities are extracted as described in the subsequent sections. The
combined feature set is given as an input to a probabilistic neural network for the
classification. Next the relevant features are selected using a forward feature
selection algorithm. The performance is analyzed using a $10$-fold cross
validation. Finally the classification performance is compared with that of a kernel
based support vector machines. Next we describe the details of the feature
extraction algorithms.


\subsection{Feature Extraction}\label{FeatExt}
\subsubsection{Time Domain Statistics (TDS)}
The first subset of features is computed from the sample statistics for each
segment within a pattern \cite{hudgins1993new}. The mean value for the
$\vj$-th segment (see eq. (\ref{Segmentation})) is defined as
\begin{equation}
f_t(\vj) =   \frac{1}{N_w}\sum_{l=1}^{N_w} s_{\vj}(l)
\end{equation}

\pagebreak

\twocolumn

Similarly, the variance, skewness and kurtosis are computed. Finally, for a $p$-th
pattern, the $4$ features from each segment and $M$ channels are grouped into
the TDS feature vector.

\subsubsection{Inter-channel statistics (ICS)}
This subset of features is based on maximum cross-correlation
\cite{smith2013comparison} among the corresponding $w$-th segments of the
two channels $i$ and $j$ of the $p$-th signal and is defined as
\begin{equation} \label{fsetI01}
\rho_{w,p}^{i,j} =   max \bigg({E}\big \{ s_{i,w,p}(l) s_{j,w,p}(l)\big\}\bigg)
\end{equation}
Note that the cross-correlation function need not have a maxima at the zero-lag.
Here the values of $i$ and $j$ are chosen from the set $C_1$ of ${M}\choose{2}$
ordered pairs defined below.
 \begin{eqnarray}
 C_1 = \{(i,j): i = 1,2,\cdots,M-1;   \nonumber \\
  j = 2,3,\cdots,M ~~\text{and}~~ i \neq j \}
 \end{eqnarray}

\subsubsection{Log Moments of Fourier Spectra (LMF)}
The logarithms of moments and their ratios from the frequency domain are
computed for the EMG segments based on \cite{Al-Timemy2013}. Consider the
L-point Fourier transform of a segment of the $m$-th channel of the  $p$-th EMG
signal
\begin{equation}\label{DFT}
S_{\vj}(k) = \sum_{l=1}^{L} s_{\vj}(l) \exp \bigg(-\frac{j 2 \pi lk}{L}\bigg )
~~~~~ k = 1,\cdots, L
\end{equation}
Now the squared magnitude spectrum is defined as
\begin{equation} \label{psd}
\psi_{\vj}(k) = |S_{\vj}(k)|^2
\end{equation}
The $i$-th frequency domain moment from the Fourier transform is defined as
\cite{Al-Timemy2016}
\begin{eqnarray} \label{momentSet1}
g_{\vj}(i) &=&  \sqrt{\sum_{k=1}^{L} k^i \psi_{\vj}(k)} \nonumber \\
\end{eqnarray}
where $i \in \{0,\cdots,6\}$. Now, the moment features are defined as
\begin{eqnarray} \label{fset1}
f_{\vj}(1) &=& \ln g_{\vj}(0) \nonumber \\
f_{\vj}(2) &=& \ln g_{\vj}(2) \nonumber \\
f_{\vj}(3) &=& \ln g_{\vj}(4) \nonumber \\
f_{\vj}(4) &=&  \ln {g_{\vj}(0)} -\frac{1}{2} \ln \big(g_{\vj}(0)- g_{\vj}(2)\big)
\nonumber  \\ &&  -\frac{1}{2} \ln(g_{\vj}\big(0)-g_{\vj}(4)\big)   \\
f_{\vj}(5)&=& \ln {g_{\vj}(2)} -\frac{1}{2} \ln \big(g_{\vj}(0) g_{\vj}(4)\big)
\nonumber  \\
f_{\vj}(6) &=& \ln {g_{\vj}(0)} -\frac{1}{4} \ln \big(g_{\vj}(1) g_{\vj}(3)\big)
\nonumber  \\
f_{\vj}(7) &=& \ln {g_{\vj}(0)} -\frac{1}{4} \ln \big(g_{\vj}(2) g_{\vj}(6)\big)
\nonumber
\end{eqnarray}

The pair wise features based on moment products are
\begin{equation} \label{fset2}
 f_{\vj}(n) =  \frac{1}{2}\ln {g_{\vj}(i)g_{\vj}(j)}   ~ \text{with} ~ n = 8, \cdots,
17
 \end{equation}
where the values of $i$ and $j$ are chosen from the set $C_2$ of $10$ ordered
pairs defined below.
 \begin{equation}
 C_2 = \{(i,j): i = 1,2,3,4; j = 2,3,4,5 ~~\text{and}~~ i \neq j\}
 \end{equation}
 Note that the features $f_{\vj}(n) $ for $n = 6, \cdots, 17$ are newly proposed.

\subsubsection{Spectral Band Powers (SBP) }
The spectral features were previously proposed for the EMG pattern recognition
in \cite{khushaba2014towards}. In this work, the spectral band power features
are extracted as follows. For each channel of the $p$-th pattern we compute the
auto-regression model coefficients $\{a_{i}\}$  assuming a model of order $\nu $
using the Burg's method described in \cite{StoicaMoses2005}.
\begin{equation}
H(\omega_k) = \frac{1}{A(\omega_k)}
\end{equation}
here $A(\omega)$ is derived from the z-transform $A(z) =  \sum_{i=0}^{\nu} a_i
z^i$. Next the power spectral density estimate is given by
\begin{equation}
\Psi(\omega_k) =  |H(\omega_k)|^2 \sigma_{burg}^2
\end{equation}
here $\sigma_{burg}^2$ is the error variance computed in the Burg's method.
Finally the SBP features are evaluated by dividing the spectrum into $N_b$ bands
and computing the respective powers in those bands. For $k = 1, \cdots, K_1;
K_1+1, \cdots, K_2; K_2+1, \cdots, K_3; \cdots$, the power within a band is
\begin{equation}  \label{SBP01}
\eta_b =  \sum_{k \in b^{th} band} \Psi(\omega_k)  ~~~~  b = 1, \cdots, N_b
\end{equation}

\subsubsection{Local Binary Patterns}
For the $w$-th segment of the $m$-th channel in the $p$-th pattern, the local
binary patterns (LBP) can be computed as follows
\begin{equation}
s_{lbp}(i) = \sum_{j} \Phi_j^{b(i)}
\end{equation}
here, $\Phi_j$ are the elements of the basis vector used in computing the LBPs for
the $i$-th interval. Next the exponents $b(i)$ are given by
\begin{equation}
 b_i = u \big( \{ g_i -  g_c \} \big)
\end{equation}
here $u(\cdot)$ is the step function and  $g_i$ are the values of the signal over a
sliding window of length $N_{LBP}$ and $g_c$ is the mean of a subset of the
same vector, for details refer to \cite{maenpaa2003local}. Finally, the features are
computed by counting the number of  LBP values above and below a certain
threshold $\tau_{lbp}.$
\begin{eqnarray} \label{feat_lbp}
 f^m_{lbp}(1) =  \# \{s_{lbp} \leq \tau_{lbp} \}   \nonumber \\
 f^m_{lbp}(2) =  \#\{  s_{lbp}  >   \tau_{lbp} \}
\end{eqnarray}
here $\# \{ A\}$ represents the cardinality of a set $A$.

\subsection{Learning using PNN}
The application of probabilistic neural networks (PNN) for sEMG  signal
classification is discussed in \cite{bu2009hybrid}. The learning methodology for
multi-class problem using the PNN  is based on learning the multi-modal
multi-variate PDF of the data using the Parzen's method. Ideally, each category
has a corresponding mode in the multi-modal pdf. For the test data, the posterior
probabilities that a pattern belongs to each of these modes are computed. Next
based on maximum among these posterior probabilities, the corresponding class
is assigned to the test pattern.  The key feature of the PNNs is that there is a
separate neuron for each training pattern and hence the size of the network
depends on the size of the training data  \cite{Specht1990}.

\section{Implementation and Results}
\label{sec:typestyle}
\subsection{Physical Action Dataset}
The data set of interest is taken from the UCI Machine learning repository
\cite{Lichman:2013}. The dataset consists of the EMG signals recorded using the
Delsys EMG electrodes on $S = 4$ subject while they performed $C = 20$
different physical activities of which $10$ were aggressive and $10$ were normal
activities as listed in table \ref{Categories}. Each subject repeats the physical
action $R=15$ times. There were $8$ EMG electrodes placed on each subject,
$4$ on biceps and triceps muscle groups of the upper limbs and $4$ on the thighs
and hamstring muscle groups of the lower limbs. Finally, each channel  consists of
approximately $10,000$ values. Hence, in this study the total sample size
becomes $P = 4 \times 20 \times 15 = 1200$.

\begin{table}[htb]
\begin{center}
\caption{Physical action categories and class labels} \label{Categories}
\begin{tabular}{c  c  c  c  }
 \hline
Label  & Normal & Label  & Aggressive   \\
 \hline
1 & Bowing   &    11   &   Elbowing \\
2 & Clapping  &    12  &   Frontkicking\\
 3 & Handshaking  &    13  &   Hammering  \\
 4 & Hugging  &    14  &     Headering\\
 5 & Jumping  &    15  &  Kneeing   \\
 6 & Running  &    16  &    Pulling \\
 7 & Seating  &    17 &   Punching  \\
 8 & Standing  &    18  &  Pushing   \\
 9 & Walking  &    19  &  Side-kicking   \\
 10 &Waving  &    20  &  Slapping   \\
 \hline
\end{tabular}
\end{center}
\end{table}
\subsection{Application of PNN based classification}
The features from $M = 8$ channels for each modality are computed as follows.
Based on $R= 15$ and total signal length $=10000$, the EMG signal length per
trial is $N= 666$ . The number segments is $N_s = 1$, and the width $N_w = N$,
Next, the feature subsets from the different feature extraction modalities
mentioned in section (\ref{FeatExt}) are computed and combined into a full
feature vector.
\begin{eqnarray}
\vf_{all}(p) &=& \big[\vf_{TDS}(p), ~  \vf_{ICS}(p),~  \vf_{LMF}(p),  \nonumber \\
&&
~ \vf_{SBP}(p),~ \vf_{LBP}(p) \big]
\end{eqnarray}
As presented in the table \ref{subsetCard}, the cardinality of each subset (except
ICS) is $8 \times c_{mod}$, where $c_{mod}$ is the cardinality of the feature
subset per modality per channel. For example, for time domain features, the
cardinality is $8 \times 4 = 32$.

\begin{table}[htb]
\begin{center}
\caption{Cardinalities of each feature subset through combination of the 8 channels}\label{subsetCard}
\begin{tabular}{c  c  c  c  c  c }
 \hline
 Feature Subset  & TDS & ICS  & LMF & SBP & LBP  \\
 \hline
No. of features  & 32  &    12   & 136  & 80 &  16 \\
 \hline
\end{tabular}
\end{center}
\end{table}
Now, based on the pairs given in table \ref{Tab-ICSPairs}, the cross-correlation
features ($6$ each) are computed for the channels from the upper limbs and the
lower limbs respectively. The LMF caridnality is based on the  fact  that each
channel has $17$ LMF features amounting to a total of $136$ features. For the
spectral band powers, the number of bands is chosen to be $N_b = 10$ leading to
a total of $80$. Finally, the $2$ LBP features are computed per channel counting
to a total of $12$.
\begin{table}[htb]
\begin{center}
\caption{Channel pairs considered for computing the cross-correlation features}\label{Tab-ICSPairs}
\begin{tabular}{c  c  c  c }
 \hline
 Index & Upper Limbs & Index  & Lower Limbs  \\
 \hline
1& (3,4)  & 7& (7,8) \\
2& (2,4)  &  8& (6,8)  \\
3& (2,3) &  9& (6,7)  \\
4& (1,4) &  10& (5,8) \\
5& (1,3) &  11&(4,7) \\
6& (1,2) & 12& (5,6) \\
 \hline
\end{tabular}
\end{center}
\end{table}

\subsection{Analysis}
\label{sec:majhead} In the cross-validation strategy, the sequence of the feature
vectors and the output labels is shuffled to avoid any learning bias in the
classification. Next the algorithm performance is evaluated by averaging the $10$
fold cross-validation based classification accuracies $\alpha $ and kappa
accuracies $\kappa$ across $M_c = 100$ Monte-Carlo runs.

\subsubsection{Sequential Forward Feature Selection} Now we apply the forward
feature selection process to the feature set $\vf_{all}$ as follows. The process
begins by selecting a feature that has the best accuracy among the individual
features. This feature is placed in $\vf_{sel}$, the selected feature subset. In the
next step, another performance maximizing feature is selected and added to the
same subset $\vf_{sel}$ and this process is repeated until the classification
performance converges to a maximum. The list of selected features is presented
below. The feature index in table \ref{SelFeatIndex1} is based on the cardinality of
the subsets given in table \ref{subsetCard}.

\begin{table}[htb]
\centering
\caption{Selected features (index) from each modality taken from the 8
channels that have significant contribution to the classification performance}
\label{SelFeatIndex1}
\begin{tabular}{c  c  c  c  c  c }
 \hline
 Subsets & TDS & ICS & LMF & SBP & LBP \\
\hline
Channels &  & & && \\
1  & 1  & 6 & 4, 10, 11, 14, 16 & - & - \\
2  & 7 & 3, 6 & 24, 29, 30, 34 & 11, 13  & -\\
3 & 9,10, 11 & 3 & 41, 42, 48 & 23, 26, 30 & 5   \\
4&   -   & - & - & 31 &- \\
5 & 17  & 12 & 78 & - & 9\\
6 & 23 & 12 & - & - & - \\
7& 25  & - & 107 & 66 & - \\
8 & 29 & - & 126 & - & -  \\
 \hline
\end{tabular}

\end{table}

Among the time domain statistics, the mean value  of the EMG signal is the key
feature selected from the channels $1, 3, 5, 7, 8$. Next the variance statistic from
the channel $3$ is selected. Followed by the skewness from the channels $2, 3$
and $6$. Finally the kurtosis is not selected and hence irrelevant for improving
the classification accuracy.

\begin{table}[htb]
\centering
\caption{Channel pairs that have significant cross-correlation and contribute to physical action classification}
\label{Tab-ICS}
\begin{tabular}{c  c  }
 \hline
Upper Limbs  & Lower Limbs \\
\hline
 (1,2); (2,3)  &   (5,6)  \\
 \hline
\end{tabular}
\end{table}

Next in terms of the inter channel statistics, the correlation between the EMG
signals of the channels $(1,2)$  (i.e., the right biceps and triceps) and $(2,3)$ (i.e.,
the right triceps and left biceps), and again the correlation between the channels
$(5,6)$ (i.e., the right thigh and hamstring) have been found to be contributing to
the classification (see table \ref{Tab-ICS}).
 Next among the FFT based moment ratios/products (LMF),  the relevant features in each
channel are specified in the table \ref{SelFeatIndex1}.  Clearly the moment ratio
defined as $f_{\vj}(7) $ given in eq. \ref{fset1} is the most relevant feature among
the LMF features. Next in the channels $1,2$ the moment ratios defined in eq.
\ref{fset2} are found to be of relevance. Next the moment product features
$f_{\vj}(10)$ in the eq. \ref{fset2} from the channel $5$ and $f_{\vj}(5)$ from the
channel $7$ also contribute to the classification. Finally, among the LMF features,
those from the channels $4$ and $6$ are found to be irrelevant. From the feature
selection we note that only  the average powers from the bands $2, 3, 4$ and $7$
as mentioned in table \ref{SelFeatIndex1} are found to be relevant. Finally, in the
case of the local binary patterns based features, the features $f_{lbp}(1)$  from
the channels $3, 5$  and $7$ are contributing to the classification.

 \subsubsection{Channel Relevance Analysis}
\begin{figure}
  \centering
  \includegraphics[width=7cm]{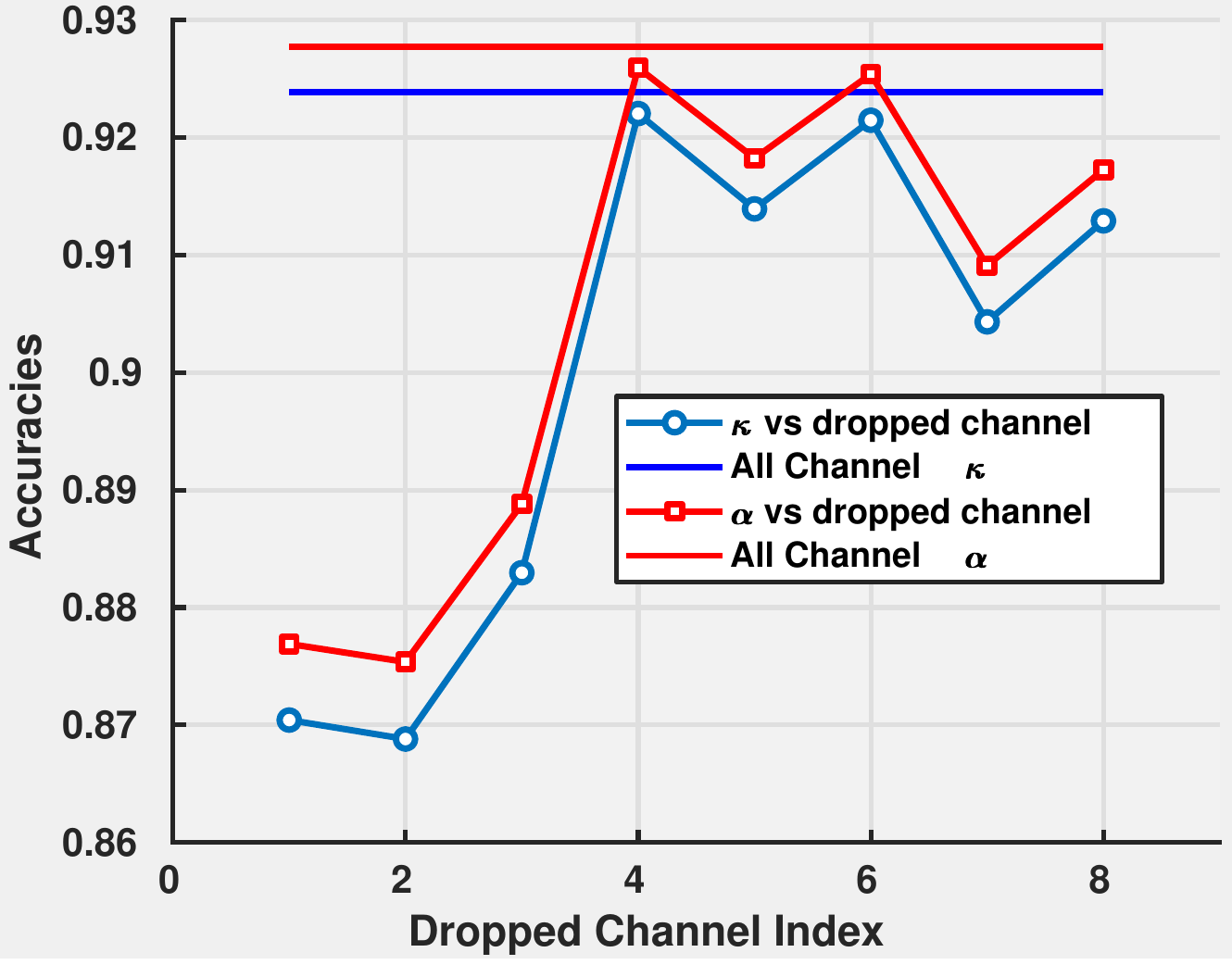}\\
  \caption{Sensitivty of classification performance to EMG channels}\label{fig_chsens}
\end{figure}
The relevance of the channels toward classification is measured in terms of
$\alpha$ and $\kappa$. In this approach, we perform the classification of the
EMG data, omitting one channel at a time from the selected features as given in
table \ref{SelFeatIndex1}, to compute the $\alpha$ and $\kappa$. From the
$M_c= 100$ runs of each of the $8$ cases, we obtain the average classifier
performance in a $10-$fold cross validation framework and present it in fig.
\ref{fig_chsens}. From this figure, it is clear that the channels $1,2$ and $3$ i.e.,
the features of the EMG data from the electrodes placed on the right biceps, right
triceps and left biceps has the most relevance. Next features corresponding to the
channels $7$ i.e., left thigh can be ranked second. Finally, the channels $4, 5, 6$
and $8$ (left triceps, right thigh, right hamstring and left hamstring) fall third in
terms of relevance toward classification.

\subsubsection{Classification performance and comparisons}

\begin{figure}

%
\begin{minipage}[b]{1.0\linewidth}
  \centering
  \centerline{\includegraphics[width=8cm]{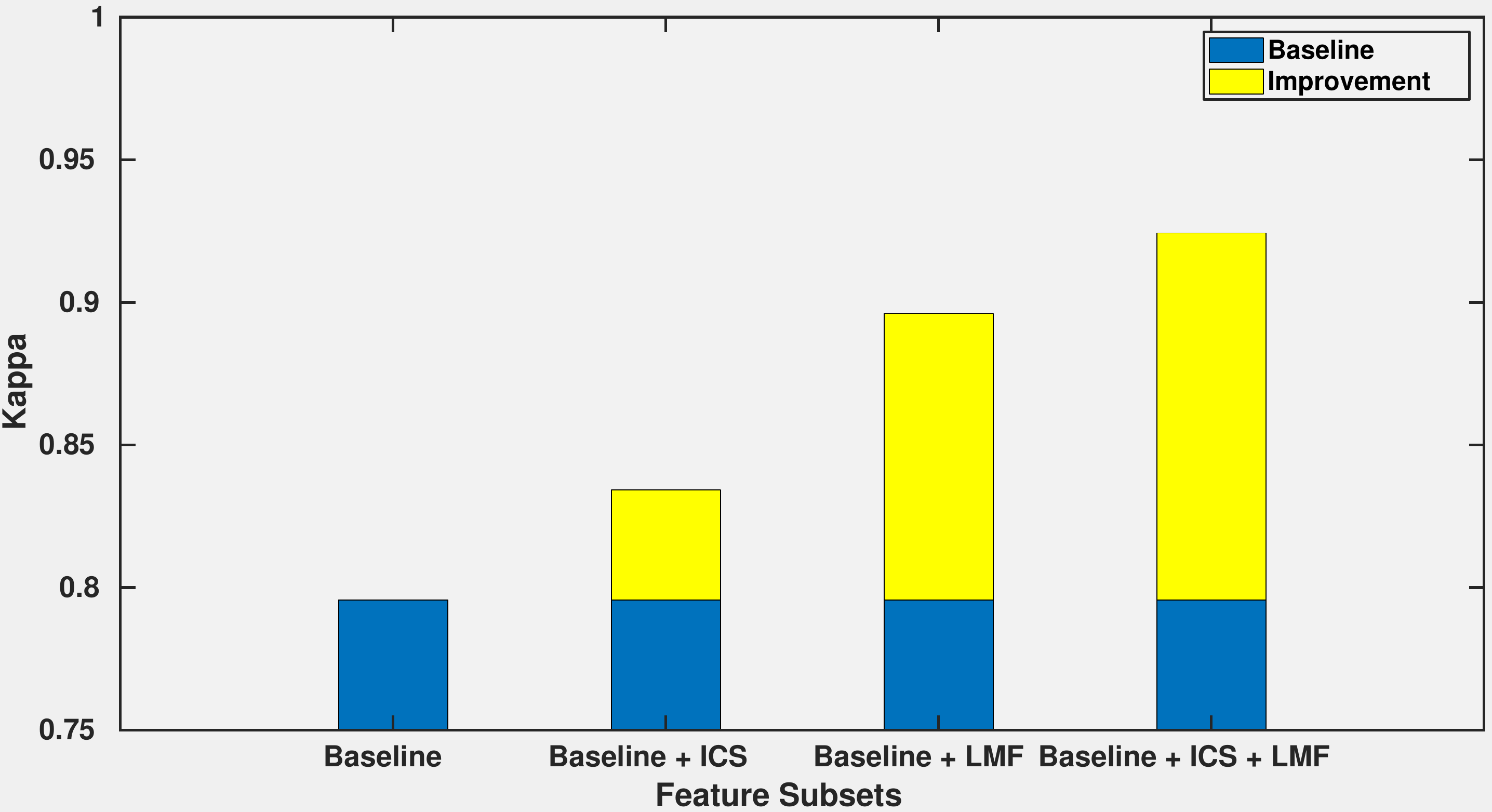}}
  \centerline{(b)Cohen's Kappa Coefficient $\kappa$}\medskip
\end{minipage}
   \caption{The performance improvement in $\kappa$ due to inclusion of the
proposed and selected features}\label{accfig}
\end{figure}

The average classification performance when all the feature subsets are used is
$\alpha = 93\%$ and $\kappa = 0.925$. Based on only the selected features using
the PNN algorithm the average classification accuracy  is $\alpha = 92.75\%$ and
kappa accuracy $\kappa = 0.924$. The selected features included a few features
from each subset along with the newly proposed spectral moment ratio and
products and the inter channel correlation features. In the fig. \ref{accfig}, the
term \textit{Baseline} refers to the subset of $22 $ selected features that are
previously used in existing literature. Next, \textit{ICS} refers to the selected
inter-channel correlation features (see table \ref{Tab-ICS}) and \textit{LMF} refers
to the proposed spectral moment ratios and products (see tables
\ref{SelFeatIndex1}). As shown in fig \ref{accfig} , the inclusion of the \textit{ ICS}
features improves the performance by $4\%$, the \textit{LMF} features improves
the kappa performance by $9\%$ and overall improvement due to the proposed
features is $11.75\%$.

 The overall performance is given in the confusion matrix in table \ref{Tab-CM1}.
Recall that the physical actions have two broad categories, the normal and the
aggressive.  From the confusion matrix in table \ref{Tab-CM1}, the first quadrant
of the matrix has a significantly stronger diagonal compared to the fourth
quadrant which suggests that the PNN classifier is highly successful in learning the
normal actions compared to the aggressive actions. The second and the third
quadrants are nearly all zeros suggesting that there is very little confusion
between the normal and the aggressive activities. In the fourth quadrant, the
section of the confusion matrix corresponding to the aggressive activities, shows
that there is a slightly higher number of miss-classifications. For example, $16$
patterns from the class $19$ get classified as the class $12$ which means there is
a considerable confusion between the front kicking and the side kicking actions.
Next, the classification performance of the PNN is compared with that of the
multi-class SVM with different kernels. The multi-class SVM is implemented with a
$C(C-1)/2$ binary SVM tree using the polynomial kernel. The kappa accuracy
$\kappa =0.91$ and the classification accuracy  $\alpha = 91.5\%$. The SVM is
implemented with the same set of selected features from the SFS algorithm used
in the PNN based approach. A similar feature analysis also applies to the SVM
based classification of the physical actions. Finally, there is
similar confusion between the front kicking and the side kicking
activities, again see the table \ref{Tab-CM2}.

\onecolumn
\begin{table}[h!]
\begin{center}
\caption{Confusion Matrix from PNN with 10-fold cross validation}
\label{Tab-CM1}
\begin{tabular}{c  c   c   c   c  c  c  c c c c c c c c c c c c c c }
 \hline
Predicted  &  1 & 2 &3 &4&5&6&7&8&9&10&11&12&13&14&15&16&17&18&19&20\\
\hline
True  \\
1& 58 &0 & 2&0&0&0&0&0&0&0&0&0&0&0&0&0&0&0&0&0\\
2& 0 &60 & 0&0&0&0&0&0&0&0&0&0&0&0&0&0&0&0&0&0\\
3& 0 & 1 & 58& 0&0&0&0&0&0&1&0&0&0&0&0&0&0&0&0&0\\
4& 0 &2 &0& 58& 0&0&0&0&0&0&0&0&0&0&0&0&0&0&0&0\\
5& 0 &0 &0&0&55&5&0&0&0&0&0&0&0&0&0&0&0&0&0&0\\
6& 0 &0 &0&0& 1&59&0&0&0&0&0&0&0&0&0&0&0&0&0&0\\
7& 0 &0 &0&0& 0&0&59&1&0&0&0&0&0&0&0&0&0&0&0&0\\
8& 0 &0 &0&0& 0&0&3&57&0&0&0&0&0&0&0&0&0&0&0&0\\
9& 0 &0 &0&0& 0&0&0&0&60&0&0&0&0&0&0&0&0&0&0&0\\
10& 0 &1 &0&0& 0&0&0&0&0&59&0&0&0&0&0&0&0&0&0&0\\
11& 0 &0 &0&0& 0&0&0&0&0&0&49&1&2&1&3&0&2&0&0&2\\
12& 0 &0 &0&0& 0&0&0&0&0&0&0&52&2&0&0&0&0&0&4&2\\
13& 0 &0 &0&0& 0&0&0&0&0&0&0&1&56&0&0&0&3&0&0&0\\
14&0 &0 &0&0& 0&0&0&0&0&0&0&1&0&52&0&0&0&0&0&7\\
15& 0 &0 &0&0&0&0&0&0&0&0&1&0&0&0&55&2&0&2&0&0\\
16& 0 &0 &0&0& 0&0&0&0&0&0&1&0&0&0&0&56&1&1&0&1\\
17& 0 &0 &0&0& 0&0&0&0&0&0&0&0&1&0&0&0&59&0&0&0\\
18& 0 &0 &0&0& 0&1&0&0&0&0&0&0&0&0&2&4&1&53&0&0\\
19& 0 &0 &0&0& 0&0&0&0&0&0&1&16&0&1&0&0&0&0&41&1\\
20& 0 &0 &0&0& 0&0&0&0&0&0&0&1&0&2&0&0&0&0&0&57\\
 \hline
\end{tabular}
\end{center}
\end{table}

\begin{table}[htb]
\centering
\caption{Confusion Matrix from multi-class SVM with 10-fold cross
validation}\label{Tab-CM2}
\begin{tabular}{c  c   c   c   c  c  c  c c c c c c c c c c c c c c }
 \hline
Predicted  &  1 & 2 &3 &4&5&6&7&8&9&10&11&12&13&14&15&16&17&18&19&20\\
\hline
True  \\
1 & 59 &0 & 1&0&0&0&0&0&0&0&0&0&0&0&0&0&0&0&0&0\\
2\vline & 0 &58 & 0&2&0&0&0&0&0&0&0&0&0&0&0&0&0&0&0&0\\
3& 1 & 0 & 58& 0&0&0&0&0&0&1&0&0&0&0&0&0&0&0&0&0\\
4& 0 &1 &0& 59& 0&0&0&0&0&0&0&0&0&0&0&0&0&0&0&0\\
5& 0 &0 &0&0& 58&2&0&0&0&0&0&0&0&0&0&0&0&0&0&0\\
6& 0 &0 &0&0& 3&57&0&0&0&0&0&0&0&0&0&0&0&0&0&0\\
7& 0 &0 &0&0& 0&0&57&3&0&0&0&0&0&0&0&0&0&0&0&0\\
8& 0 &0 &0&0& 0&0&0&59&0&1&0&0&0&0&0&0&0&0&0&0\\
9& 2 &0 &0&0& 0&2&0&0&58&0&0&0&0&0&0&0&0&0&0&0\\
10& 1 &1 &0&0& 0&0&0&0&0&58&0&0&0&0&0&0&0&0&0&0\\
11& 0 &0 &0&0& 0&0&0&0&0&0&57&0&1&1&1&0&0&0&0&0\\
12& 0 &0 &0&0& 0&0&0&0&0&0&0&47&2&0&0&0&0&0&11&0\\
13& 0 &0 &0&0& 0&0&0&0&0&0&3&0&52&0&0&0&3&2&0&0\\
14& 0 &0 &0&0& 0&0&0&0&0&0&0&1&0&53&0&0&0&0&2&4\\
15& 0 &0 &0&0& 0&0&0&0&0&0&7&4&0&0&48&1&0&0&0&0\\
16& 0 &0 &0&0& 0&0&0&0&0&0&1&0&0&0&1&54&1&2&1&0\\
17& 0 &0 &0&0& 0&0&0&0&0&0&1&0&2&1&0&0&55&1&0&0\\
18& 0 &0 &0&0& 0&0&0&0&0&0&1&0&0&1&1&4&0&49&1&3\\
19& 0 &0 &0&0& 0&0&0&0&0&0&0&9&0&1&0&0&0&0&50&0\\
20& 0 &0 &0&0& 0&0&0&0&0&0&0&0&0&7&0&0&0&0&1&52\\
 \hline
\end{tabular}
\end{table}
\twocolumn

\section{Conclusion}
\label{sec:illust} In this study, we have implemented a multi-category
classification framework based on  the probabilistic neural networks to categorize
the physical actions using the features derived from eight channels of the surface
EMG data. A set of $276$ features were extracted from various modalities
including the statistical features in time domain, the inter channel cross
correlation features, logarithms of moment ratios of the Fourier spectra, the
mean band powers of power spectral density estimates based on the Burg
algorithm, and the features based on the local binary  patterns. Using the
sequential forward selection algorithm a set of $37$ features from above $276$
features were found to be relevant for the physical action classification.
 The selected features included a few features from each subset and also
a set of modified spectral moment products/ratios and the inter channel
correlation features. Based on the selected features using the PNN algorithm, the
average classification accuracy  is $92.75\%$ and kappa accuracy is $0.924$. In
terms of channel relevance, the EMG data from the upper limbs has greater
significance in physical action classification. Next the classification performance is
similar to that of the multi-class SVM (classification $91.5\%$ and the kappa
accuracy of $0.91$ which requires an extra computational time. The future plans
for this research have two important directions. The first is to explore the
information theoretic learning methods to improve the classification accuracy and
identify a set of relevant features. A second direction is to conduct actual
experiments to acquire the EMG measurements for controlling an orthotic
exoskeleton for an upper limb. Finally, the overall goal is to find an optimal
combination of learning algorithms and control strategies for the upper limb
exoskeletons.


\bibliographystyle{IEEEtran}
\bibliography{PA_REF}

\end{document}